\documentclass[12pt,preprint]{emulateapj}
\usepackage{graphicx}

\shorttitle{The Most Eccentric Exoplanet Known}
\shortauthors{Stephen R. Kane et al.}
\slugcomment{Submitted for publication in the Astrophysical Journal}

\begin{document}

\title{Evidence for Reflected Light from the Most Eccentric Exoplanet
  Known}

\author{
  Stephen R. Kane\altaffilmark{1},
  Robert A. Wittenmyer\altaffilmark{2,3,4},
  Natalie R. Hinkel\altaffilmark{1,5},
  Arpita Roy\altaffilmark{6,7},
  Suvrath Mahadevan\altaffilmark{6,7},
  Diana Dragomir\altaffilmark{8},
  Jaymie M. Matthews\altaffilmark{9},
  Gregory W. Henry\altaffilmark{10},
  Abhijit Chakraborty\altaffilmark{11},
  Tabetha S. Boyajian\altaffilmark{12},
  Jason T. Wright\altaffilmark{6,7},
  David R. Ciardi\altaffilmark{13},
  Debra A. Fischer\altaffilmark{12},
  R. Paul Butler\altaffilmark{14},
  C.G. Tinney\altaffilmark{2,3},
  Brad D. Carter\altaffilmark{4},
  Hugh R.A. Jones\altaffilmark{15},
  Jeremy Bailey\altaffilmark{2,3},
  Simon J. O'Toole\altaffilmark{16}
}
\email{skane@sfsu.edu}
\altaffiltext{1}{Department of Physics \& Astronomy, San Francisco
  State University, 1600 Holloway Avenue, San Francisco, CA 94132,
  USA}
\altaffiltext{2}{School of Physics, University of New South Wales,
  Sydney, NSW 2052, Australia}
\altaffiltext{3}{Australian Center for Astrobiology, University of New
  South Wales, Sydney, NSW 2052, Australia}
\altaffiltext{4}{Computational Engineering and Science Research
  Centre, University of Southern Queensland, Toowoomba, Queensland
  4350, Australia}
\altaffiltext{5}{School of Earth \& Space Exploration, Arizona State
  University, Tempe, AZ 85287, USA}
\altaffiltext{6}{Department of Astronomy and Astrophysics,
  Pennsylvania State University, 525 Davey Laboratory, University
  Park, PA 16802, USA}
\altaffiltext{7}{Center for Exoplanets \& Habitable Worlds,
  Pennsylvania State University, 525 Davey Laboratory, University
  Park, PA 16802, USA}
\altaffiltext{8}{Las Cumbres Observatory Global Telescope Network,
  6740 Cortona Drive, Suite 102, Goleta, CA 93117, USA}
\altaffiltext{9}{Department of Physics and Astronomy, University of
  British Columbia, Vancouver, BC V6T1Z1, Canada}
\altaffiltext{10}{Center of Excellence in Information Systems, Tennessee
  State University, 3500 John A. Merritt Blvd., Box 9501, Nashville,
  TN 37209, USA}
\altaffiltext{11}{Division of Astronomy, Physical Research Laboratory,
  Navrangpura, Ahmedabad 380009, India}
\altaffiltext{12}{Department of Astronomy, Yale University, New Haven,
  CT 06511, USA}
\altaffiltext{13}{NASA Exoplanet Science Institute, Caltech, MS 100-22,
  770 South Wilson Avenue, Pasadena, CA 91125, USA}
\altaffiltext{14}{Department of Terrestrial Magnetism, Carnegie
  Institution of Washington, 5241 Broad Branch Road, NW, Washington,
  DC 20015-1305, USA}
\altaffiltext{15}{University of Hertfordshire, Centre for Astrophysics
  Research, Science and Technology Research Institute, College Lane,
  AL10 9AB, Hatfield, UK}
\altaffiltext{16}{Australian Astronomical Observatory, PO Box 915,
  North Ryde, NSW 1670, Australia}


\begin{abstract}

Planets in highly eccentric orbits form a class of objects not seen
within our Solar System. The most extreme case known amongst these
objects is the planet orbiting HD~20782, with an orbital period of
597~days and an eccentricity of 0.96. Here we present new data and
analysis for this system as part of the Transit Ephemeris Refinement
and Monitoring Survey (TERMS). We obtained CHIRON spectra to perform
an independent estimation of the fundamental stellar parameters. New
radial velocities from AAT and PARAS observations during periastron
passage greatly improve our knowledge of the eccentric nature of the
orbit. The combined analysis of our Keplerian orbital and Hipparcos
astrometry show that the inclination of the planetary orbit is $>
1.22\degr$, ruling out stellar masses for the companion. Our long-term
robotic photometry show that the star is extremely stable over long
timescales. Photometric monitoring of the star during predicted
transit and periastron times using MOST rule out a transit of the
planet and reveal evidence of phase variations during
periastron. These possible photometric phase variations may be caused
by reflected light from the planet's atmosphere and the dramatic
change in star--planet separation surrounding the periastron passage.

\end{abstract}



\section{Introduction}
\label{introduction}

Exoplanetary system architectures have revealed numerous surprises
since the first exoplanets were discovered. One of the earliest
surprises was the discovery of exoplanets in highly eccentric orbits,
for which there is no analog in the Solar System. These eccentric
orbits were discovered for giant planets, such as HD~114762b
\citep{lat89,kan11c} and 70~Vir~b \citep{mar96,kan15} with
eccentricities of 0.33 and 0.40 respectively.  Since those early
discoveries, eccentric planets have presented a significant challenge
for formation theories to account for the components of planet-planet
scattering \citep{cha08,pet14} and tidal circularization
\citep{pon11}. Such planets tend to be discovered with the radial
velocity (RV) technique since the observations are able to sample the
entire Keplerian planetary orbit. Subsequent investigations of the
eccentricity distribution of planetary orbits that take {\it Kepler}
transiting exoplanet discoveries into account show that small planets
in multi-planet systems are more likely to have low eccentricities
\citep{kan12c,van15}. The discovery and characterization of eccentric
orbits is an on-going effort to understand the evolutionary history of
these fascinating systems.

A particularly eccentric exoplanet was discovered by \citet{jon06}
orbiting the star HD~20782. With a minimum mass twice that of Jupiter
and an orbital period of 597~days, the planet is typical of
high-eccentricity planets. The orbit was further revised by
\citet{oto09} and shown to have an eccentricity as high as 0.97,
making it the highest eccentricity exoplanet yet discovered. However,
data during periastron passage is difficult to obtain for such systems
since the RV variation predominantly occurs during a very small
fraction of the orbital phase. The star continued to be monitored by
the Transit Ephemeris Refinement and Monitoring Survey (TERMS) to
improve the orbital parameters of the system \citep{kan09}. Such
orbital refinement may be used to predict and observe events that
occur during particular periods of the orbit, such as planetary
transits \citep{kan08} or phase variations \citep{kan10}.

Here we present new results for the HD~20782 system, including RVs
that sample several periastron passages and establish the planet as
the most eccentric known exoplanet. Follow-up photometry from both
ground-based and space-based telescopes rule out a transit of the
planet and show evidence of phase variations due to reflected light
from the planet close to periastron passage. Section~\ref{science}
provides background information and discusses the science motivation
for studying the system. Section~\ref{stellarprop} presents analysis
of new CHIRON spectra and the resulting fundamental parameters of the
host star as well as stellar abundances. New RV data are combined with
those published in Section~\ref{orbit} and a new Keplerian orbit for
the planet is produced. Section~\ref{astrometry} describes the use of
{\it Hipparcos} astrometry to constrain the orbital inclination of the
planet. Section~\ref{transit} discusses the transit prospects for the
system and the effects of both orbital eccentricity and
inclination. Section~\ref{photometry} presents the ground-based
photometry and an estimate of the stellar rotation period. Data from
MOST are used during the transit/periastron window to rule out a
transit and also reveal the potential presence of a reflected light
signature of the planet as it passes through periastron passage. We
discuss future observing opportunities and make concluding remarks in
Section~\ref{conclusions}.


\section{Science Motivation}
\label{science}

The eccentricity distribution of exoplanets has a well-defined shape
whereby the orbits diverge from circular beyond a semi-major axis of
$\sim 0.1$~AU \citep{but06,kan13}, inside of which tidal
circularization tends to force low eccentricity \citep{gol66,pon11}.
The observed eccentricity distribution is a clear indicator of
formation processes that are dependent upon initial system
architectures, in particular planet-planet scattering. Wide binaries
may inadvertantly create a more suitable environment for the formation
of highly-eccentric planetary orbits through gravitational
perturbations from the companion star and the triggering of planetary
ejections \citep{kai13}.

\begin{figure}
  \includegraphics[angle=270,width=8.5cm]{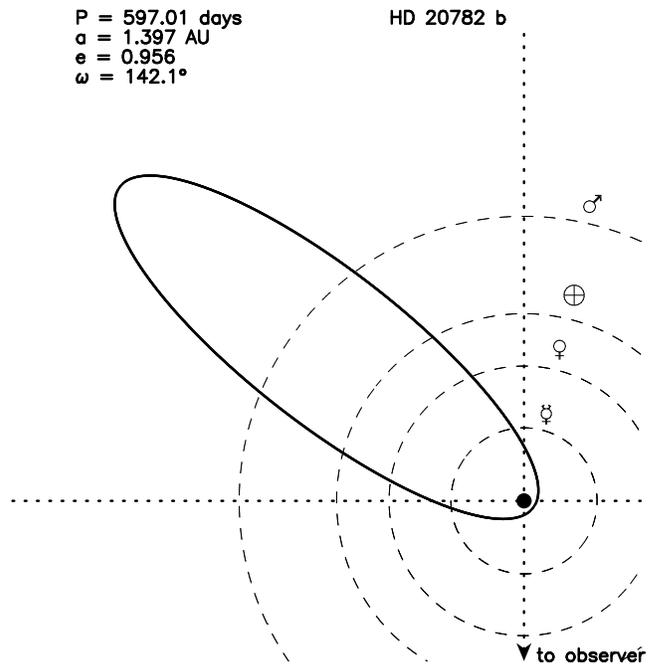}
  \caption{A top-down view of the HD~20782 system based on data
    described in this paper. The Keplerian orbit of the planet, shown
    as a solid line, is depicted using the new parameters from
    Table~\ref{planet}. The orbits of the Solar System planets (dashed
    lines) are shown for comparison.}
  \label{orbitfig}
\end{figure}

HD~20782 is part of a wide binary with HD~20781 having a projected
separation of 9,000~AU, recently described by \citet{mac14}. The known
planet orbiting HD~20782 lies at the very top of the exoplanet
eccentricity distribution, though RV measurements during the crucial
periastron passage were relatively rare. The extreme nature of the
planet's orbital eccentricity may be seen in Figure~\ref{orbitfig},
where the orbit is described using our expanded dataset (see
Section~\ref{orbit}).

Our further investigations of this system are primarily motivated by a
better characterization of the planetary orbit and performing
follow-up observations at key orbital phases that can help to
understand the nature of the planet. It is also important to establish
that the secondary object is indeed a planet since a face-on orbital
orientation would make it consistent with the eccentricity
distribution of spectroscopic binaries \citep{mei05,maz08}.

The orbital orientation depicted in Figure~\ref{orbitfig} shows that
the star--planet separation along the line of sight to the observer is
quite small, despite the $\sim$18 month orbital period. This yields a
relatively high transit probability equivalent to that of a hot
Jupiter (see Section~\ref{transit}). Thus a primary motivation for
follow-up observations is the possible detection of a planetary
transit for a long-period eccentric planet \citep{kan08}. A previous
example of such a system can be seen in the case of HD~80606b
\citep{nae01}, where the secondary eclipse of the 0.93 eccentricity
planet was detected by \citet{lau09} and later confirmed to also
exhibit a primary transit \citep{fos09,gar09,mou09}. An additional
motivation for obtaining high-precision photometry during the transit
window and periastron passage for HD~20782b is the possibility of
detecting reflected light from the planet since the small star--planet
separation will greatly increase the amplitude of the phase signature
\citep{kan10}. Such a detection would allow an estimate of the
geometric albedo of the planet and place constraints upon the
atmospheric properties and the atmosphere's radiative and advective
time scales \citep{sea05,for08}. Note that since the orbital period is
18 months, an observing opportunity for a particular point in the
orbit will only arise every 3 years since the star will be largely
inaccessible to ground-based observers for each alternate orbit.


\section{Stellar Properties}
\label{stellarprop}

A critical step in quantifying the properties of the planet lies in
understanding the host star. Here we provide new fundamental
parameters and abundances for HD~20782.


\subsection{Fundamental Parameters}
\label{stellar:sme}

We acquired a high S/N (300 second integration) spectrum of HD~20782
on the night of July 6th, 2014. The data were acquired using CHIRON, a
fiber-fed Echelle spectrometer \citep{tok13,bre14}, installed at the
1.5m telescope at Cerro Tololo Inter-American Observatory
(CTIO). CHIRON operates at a fixed wavelength range of 415--880~nm and
a resolution of $R = 79,000$. The spectrum was modeled using the
Spectroscopy Made Easy (SME) package, described in more detail by
\citet{val96,val05}. SME uses an iterative technique that combines
model atmosphere analysis with Yonsei-Yale model isochrones
\citep{dem04} that utilize {\it Hipparcos} photometry and distances
\citep{van07a,van07b}. This approach produces a self-consistent
convergence with the measured surface gravity \citep{val09}.

The results of this analysis are shown in Table~\ref{stellar},
including values for the surface gravity $\log g$, rotational velocity
$v \sin i$, atmospheric abundance [Fe/H], effective temperature
$T_{\rm eff}$ and stellar isochrone solution (mass $M_\star$, radius
$R_\star$, and age). These parameters are consistent with previous
estimates of the stellar properties, such as those calculated by
\citet{tak07}. The revised parameters demonstrate that HD~20782 is
quite similar to the Sun, with the mass and radius being crucial
properties for the subsequent analysis of the planetary companion in
this paper.

\begin{deluxetable}{lc}
  \tablecaption{\label{stellar} Stellar Parameters}
  \tablehead{
    \colhead{Parameter} &
    \colhead{Value}
  }
  \startdata
  $V$                            & 7.4 \\
  $B-V$                          & 0.63 \\
  Distance (pc)                  & $35.5 \pm 0.8$ \\
  $T_\mathrm{eff}$ (K)           & $5798 \pm 44$ \\
  $\log g$                       & $4.36 \pm 0.06$ \\
  $v \sin i$ (km\,s$^{-1}$)      & $1.7  \pm 0.5$ \\
  $[$Fe/H$]$ (dex)               & $0.01 \pm 0.03$ \\
  $M_\star$ ($M_\odot$)          & $1.02 \pm 0.02$ \\
  $R_\star$ ($R_\odot$)          & $1.09 \pm 0.04$ \\
  Age (Gyrs)                     & $5.53 \pm 1.43$
  \enddata
\end{deluxetable}


\subsection{Abundances}
\label{stellar:abund}

Both components of the wide binary system, namely HD~20781 and
HD~20782, have had their elemental abundances measured by a number of
different authors. However, due to the difference in size and spectral
type, the abundances within HD~20782 are easier to determine.  While
half as many groups have measured HD~20781 than HD~20782, there does
remain some overlap by some, such as \citet{nev09}, \citet{del10}, and
\citet{mac14} who did a more in-depth comparison of the two stars.

Per the analysis in the Hypatia Catalog \citep{hin14}, the individual
abundances within both stars were renormalized to the \citet{lod09}
solar scale. The largest measurement discrepancy between datasets,
known as the {\it spread}, was used to better quantify the uniformity,
or lack thereof, between measurements. This technique was implemented
in the Hypatia Catalog to better understand the variation in
abundances seen when employing different reduction techniques, due to
instances where the {\it spread} between groups was larger than
associated error. For the cases where variations between groups were
small, the median value was used as the ultimate abundance
measurement.

The overall median [Fe/H] content in HD~20781 was 0.1 dex, as compared
to 0.15 dex within HD~20782, where the spread was 0.03 dex and 0.17
dex, respectively.  In other words, the groups that measured HD~20781,
while fewer in number, were in closer agreement regarding the iron
abundance than those that measured HD~20782. The [Fe/H] determinations
for both stars are disparate compared to the abundances determined by
\citet{mac14}, which are not part of the Hypatia Catalog, who measured
0.04$\pm$0.03 and -0.02$\pm$0.02, respectively. These are consistent
with our new [Fe/H] determination shown in Table~\ref{stellar}.

A wide variety of $\alpha$-elements (carbon, magnesium, silicon, and
titanium), odd-Z elements (sodium, aluminum, and scandium), and
iron-peak elements (vanadium, chromium, cobalt, and nickel) have been
measured for both stars. For all elements except for [Na/Fe], the
abundance measurements for HD~20781 and HD~20782 were found to be
consistent to within error and markedly sub-solar, or $\sim$ -0.1
dex. The [Na/Fe] content in HD~20782 was found to be $\sim 2.5$ more
than in the companion HD~20781, where [Na/Fe] = -0.09$\pm$0.06 dex and
-0.22$\pm$0.04 dex, respectively.


\section{The Keplerian Orbit of the Planet}
\label{orbit}

The highly eccentric planet orbiting HD~20782 was first reported in
\citet{jon06} and updated in \citet{oto09}. We now present a further
six years of radial velocity data from the Anglo-Australian Planet
Search (AAPS). The AAPS is one of the world's longest-running planet
searches, with more than 40 planet discoveries in its 16 years of
operation \citep[e.g.][]{but01,jon10,vog10,wit12,tin11,wit14}.
HD~20782 has been observed on 52 epochs from 1998 Aug 9 to 2013 Sep 19
(Table~\ref{rvs_aat}). Precision Doppler measurements are obtained
with the UCLES echelle spectrograph \citep{die90} at the 3.9~m
Anglo-Australian Telescope (AAT). A 1-arcsecond slit delivers a
resolving power of $R\sim$45,000. Calibration of the spectrograph
point-spread function is achieved using an iodine absorption cell
temperature-controlled at 60.0$\pm$0.1$^{\rm{o}}$C. The iodine cell
superimposes a forest of narrow absorption lines from 5000 to
6200\,\AA, allowing simultaneous calibration of instrumental drifts as
well as a precise wavelength reference \citep{val95,but96}. The result
is a precise radial velocity shift measured relative to the epoch of
the iodine-free ``template'' spectrum. AAT velocities for HD~20782
span a total of 15 years and have a mean internal uncertainty of
2.4~m\,s$^{-1}$.

Orbital fits to the AAT data allowed predictions of the next
periastron passage of the planet, estimated to be 15 January 2015. We
were able to observe HD~20782 during that passage using the Physical
Research Laboratory (PRL) optical fiber-fed high-resolution
cross-dispersed echelle spectrograph (PARAS) with the Mount Abu 1.2~m
telescope in India. The PARAS spectrograph is temperature-controlled
at 25.55$\pm$0.01$^{\rm{o}}$C in an enclosure that is
pressure-controlled at 0.10$\pm$0.03~mbar. PARAS has a resolution of
$R\sim$67,000 and obtains RV data at a spectral range of 3800 to
6900\,\AA with simultaneous wavelength calibration with a
thorium-argon (ThAr) hollow cathode lamp. The uncertainties for the
PARAS measurements were derived based on the photon noise estimation
procedure explained by \citet{bou01}. Further details of the PARAS
instrument and the data reduction are described by
\citet{cha14}. PARAS observations were made under high air mass
conditions (1.7--1.9) with no Atmospheric Dispersion Corrector
(ADC). The five PARAS observations (see Table~\ref{rvs_paras})
complete our RV dataset and bring the total number of observations to
57.

\begin{deluxetable}{ccc}
  \tablewidth{0pc}
  \tablecaption{\label{rvs_aat} HD~20782 AAT Radial Velocities}
  \tablehead{
    \colhead{Date} &
    \colhead{RV} &
    \colhead{$\sigma$} \\
    \colhead{(BJD -- 2440000)} &
    \colhead{(m\,s$^{-1}$)} &
    \colhead{(m\,s$^{-1}$)}
  }
  \startdata
11035.31946 &    21.90 &  2.33 \\
11236.93065 &    -6.51 &  3.27 \\
11527.01731 &     7.32 &  3.39 \\
11630.88241 &    29.70 &  2.72 \\
11768.30885 &    -6.64 &  2.62 \\
11828.11066 &    -7.64 &  3.00 \\
11829.27449 &    -6.64 &  3.82 \\
11856.13530 &   -10.37 &  3.55 \\
11919.00660 &    -3.62 &  2.92 \\
11919.99630 &    -1.67 &  2.85 \\
11983.89009 &     4.16 &  3.32 \\
12092.30437 &    17.84 &  2.35 \\
12127.26814 &    17.70 &  2.79 \\
12152.16308 &    23.15 &  2.50 \\
12187.15965 &    22.78 &  2.53 \\
12511.20636 &    -1.26 &  2.29 \\
12592.04809 &    17.40 &  2.30 \\
12654.96031 &    15.38 &  2.34 \\
12859.30551 &  -202.48 &  1.90 \\
12946.13833 &   -18.15 &  2.08 \\
12947.12246 &   -14.27 &  1.77 \\
13004.00143 &    -0.29 &  1.85 \\
13044.02367 &     0.76 &  2.25 \\
13045.96088 &    -0.40 &  1.93 \\
13217.28800 &     9.01 &  1.71 \\
13282.22023 &    20.57 &  1.87 \\
13398.96924 &    22.14 &  1.39 \\
13403.96059 &    30.40 &  2.56 \\
13576.30688 &    -9.14 &  1.60 \\
13632.28114 &    -7.62 &  1.59 \\
13665.18659 &     6.38 &  1.72 \\
14013.21622 &    31.23 &  1.55 \\
14040.13171 &    22.12 &  1.96 \\
14153.97010 &   -11.56 &  2.10 \\
14375.24693 &    13.32 &  1.70 \\
14544.89158 &    10.26 &  2.15 \\
14776.10092 &    -7.55 &  1.85 \\
14843.02077 &     0.09 &  1.56 \\
14899.92440 &    -0.65 &  2.07 \\
15107.24701 &    16.54 &  2.78 \\
15170.05453 &    17.31 &  2.37 \\
15204.97966 &    29.22 &  1.88 \\
15253.91188 &   -78.17 &  2.35 \\
15399.32249 &    -8.19 &  1.88 \\
15426.31459 &    -6.89 &  1.71 \\
15461.23900 &   -14.81 &  2.99 \\
15519.13309 &     8.36 &  2.00 \\
15844.13584 &  -145.90 &  6.54 \\
15845.17956 &  -185.60 &  2.28 \\
15846.13671 &  -156.28 &  2.32 \\
15964.93095 &     7.77 &  2.87 \\
16499.33740 &   -11.25 &  3.03
  \enddata
\end{deluxetable}

\begin{deluxetable}{ccc}
  \tablewidth{0pc}
  \tablecaption{\label{rvs_paras} HD~20782 PARAS Radial Velocities}
  \tablehead{
    \colhead{Date} &
    \colhead{RV} &
    \colhead{$\sigma$} \\
    \colhead{(BJD -- 2440000)} &
    \colhead{(m\,s$^{-1}$)} &
    \colhead{(m\,s$^{-1}$)}
  }
  \startdata
57036.16183 & 272.25 & 4.12 \\
57038.14436 & 127.25 & 4.09 \\
57039.13336 &  61.53 & 3.65 \\
57040.15494 & 149.14 & 3.86 \\
57042.12356 & 183.06 & 2.98
  \enddata
\end{deluxetable}

The RV data shown in Tables \ref{rvs_aat} and \ref{rvs_paras} were
used to produce a revised Keplerian orbital solution. This was
performed using the RVLIN package; a partially linearized,
least-squares fitting procedure \citep{wri09}. The uncertainties for
the orbital and associated physical parameters were estimated using
the BOOTTRAN bootstrapping routines described in \citet{wan12}. To be
sure that known instabilities of the Levenberg-Marquardt-based RVLIN
algorithm did not prevent convergence at these high eccentricities, we
reduced the number of nonlinear parameters in the problem by fixing
the eccentricity at 100 values evenly spaced between 0.93 and 0.995
and selecting the value that produced the minimum $\chi^2$ fit.

A fit to the AAT and PARAS data with their instrumental uncertainties
was unsatisfactory. The rms residuals to the PARAS data are
17~m\,s$^{-1}$, with two excursions at and after RV minimum of over
20~m\,s$^{-1}$, inconsistent with typical instrumental uncertainties
of under 6~m\,s$^{-1}$. Further, the scatter about the fit to the AAT
data is 6.1~m\,s$^{-1}$, including three excursions larger than
15~m\,s$^{-1}$ (up to 17~m\,s$^{-1}$), both significantly larger than
the quoted errors of 2.3~m\,s$^{-1}$. Given that there are only 52 AAT
points, we do not expect to see 3 points ($\sim$5\%) with deviations
of 15~m\,s$^{-1}$ from Gaussian noise unless the errors are more like
6~m\,s$^{-1}$.

Some component of the scatter about the best fit is due to intrinsic
stellar variability, and some is due to the precision of the
measurements (due to both instrumental/algorithmic imprecision and
photon noise). The stellar noise should be the same for both
instruments, meaning that the large excursions seen in the PARAS data
indicate a problem with either the fit or the PARAS data.

Close examination of points near periastron reveal that the problem
must lie with the instrumental uncertainties, not the fit. PARAS and
AAT have two measurements (each) at very similar phases (the expected
change in RV between the points in each pair is $<
10$~m\,s$^{-1}$). However, in both cases the difference in velocities
is over 20~m\,s$^{-1}$, and in different directions. The combined
measurement uncertainties of the two instruments therefore must be of
order 20~m\,s$^{-1}$. 

We attempted a second fit, but inflated both instrumental
uncertainties by adding, in quadrature, 5~m\,s$^{-1}$ and
19~m\,s$^{-1}$ to the AAT and PARAS velocities, respectively. These
inflations reflect a common stellar jitter component (likely to be
around 5~m\,s$^{-1}$) and an additional, instrument-dependent
component added in quadrature. This resulted in a much more
satisfactory fit: the residuals to the best fit for the two telescopes
have standard deviations of 5.75~m\,s$^{-1}$ and 19.85~m\,s$^{-1}$,
respectively, and $\chi^2$ values of 1.03 and 1.01,
respectively. There is still a significant outlier to the AAT fit (at
15~m\,s$^{-1}$), but at 2.5$\sigma$ (using the inflated measurement
uncertainties) this is not unexpected from 52 data points.

\begin{deluxetable*}{lcc}
  \tablecaption{\label{planet} Keplerian Orbital Model}
  \tablewidth{0pt}
  \tablehead{
    \colhead{Parameter} &
    \colhead{Value (AAT)} &
    \colhead{Value (AAT+PARAS)}
  }
  \startdata
\noalign{\vskip -3pt}
\sidehead{HD 20782 b}
~~~~$P$ (days)                     & $597.099 \pm 0.049$   & $597.065 \pm 0.043$ \\
~~~~$T_c\,^{a}$ (BJD -- 2,440,000) & $17037.788 \pm 0.145$ & $17037.794 \pm 0.100$ \\
~~~~$T_p\,^{b}$ (BJD -- 2,440,000) & $17038.510 \pm 0.108$ & $17038.458 \pm 0.094$ \\
~~~~$e$                            & $0.953 \pm 0.005$     & $0.956 \pm 0.004$ \\
~~~~$\omega$ (deg)                 & $142.2 \pm 2.2$       & $142.1 \pm 2.1$ \\
~~~~$K$ (m\,s$^{-1}$)              & $114.9 \pm 4.4$       & $116.0 \pm 4.2$ \\
~~~~$M_p$\,sin\,$i$ ($M_J$)        & $1.46 \pm 0.03$       & $1.43 \pm 0.03$ \\
~~~~$a$ (AU)                       & $1.397 \pm 0.009$     & $1.397 \pm 0.009$ \\
\sidehead{System Properties}
~~~~$\gamma$ (m\,s$^{-1}$)         & $1.95 \pm 0.82$ & $1.79 \pm 0.80$ \\
\sidehead{Measurements and Model}
~~~~$N_{\mathrm{obs}}$             & 52   & 57 \\
~~~~rms (m\,s$^{-1}$)              & 5.91 & 8.06 \\
~~~~$\chi^2_{\mathrm{red}}$        & 1.0  & 1.14
  \enddata
  \tablenotetext{a}{Time of mid-transit.}
  \tablenotetext{b}{Time of periastron passage.}
\end{deluxetable*}

\begin{figure}
  \includegraphics[width=8.2cm]{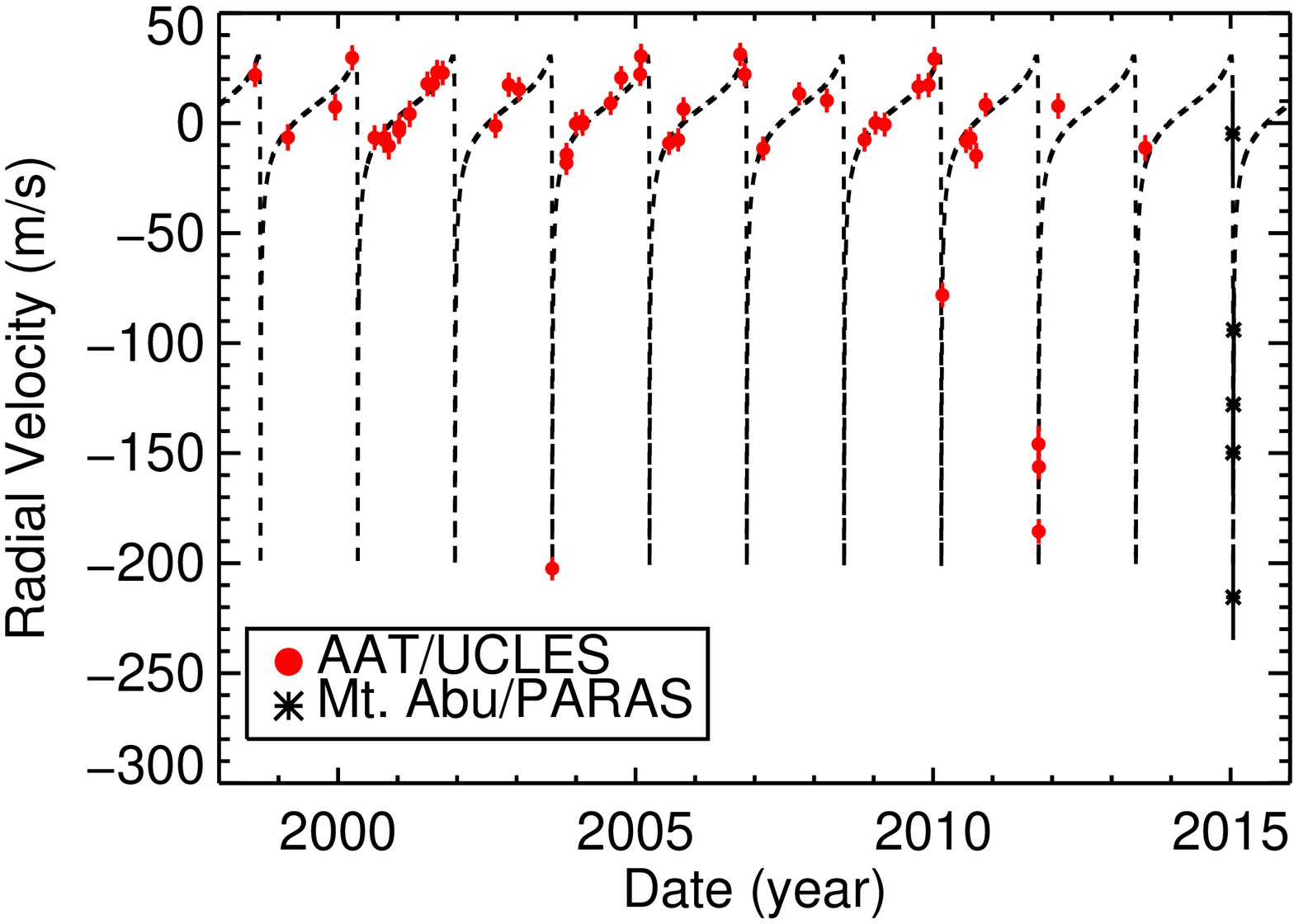} \\
  \includegraphics[width=8.3cm]{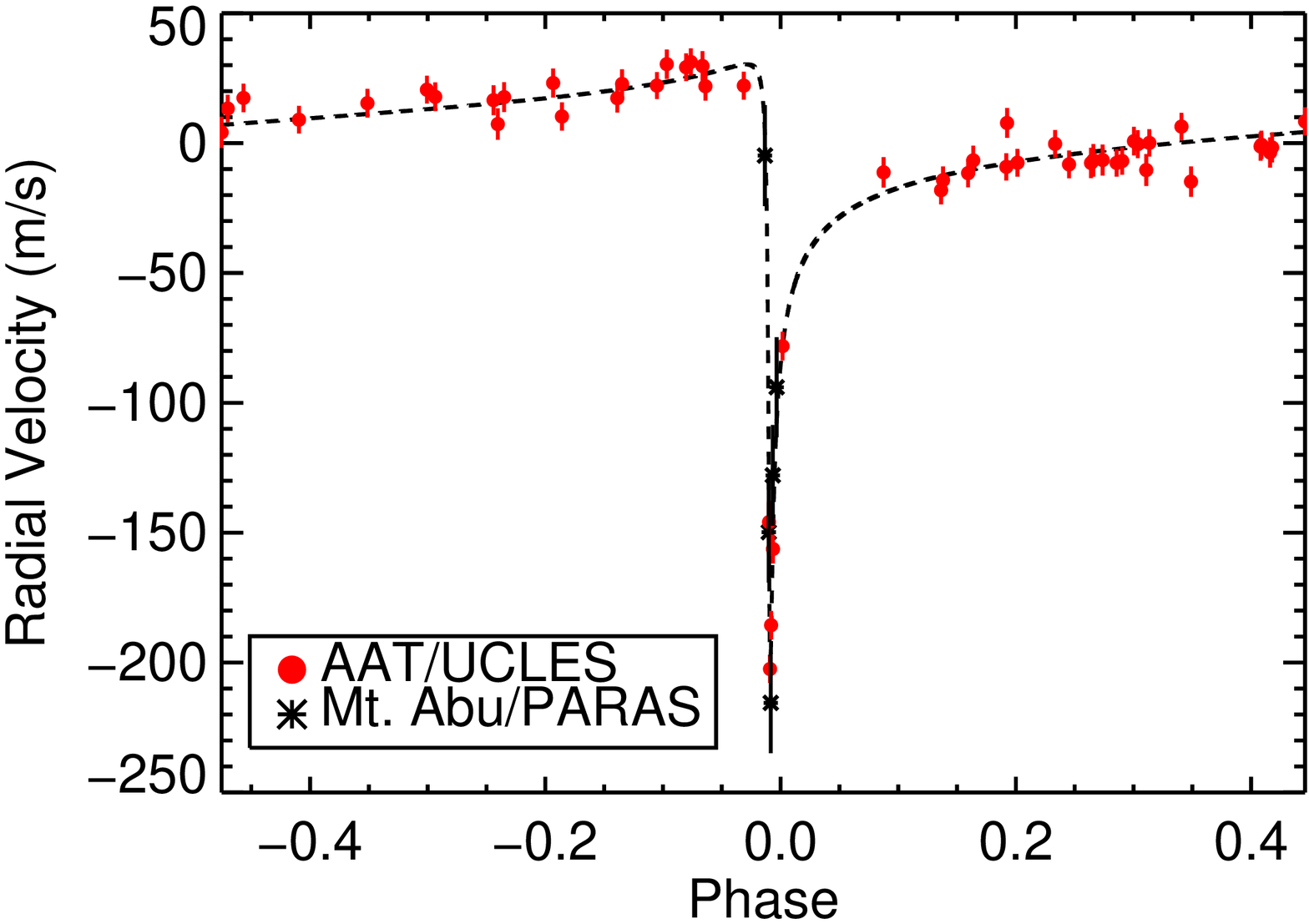} \\
  \includegraphics[width=8.4cm]{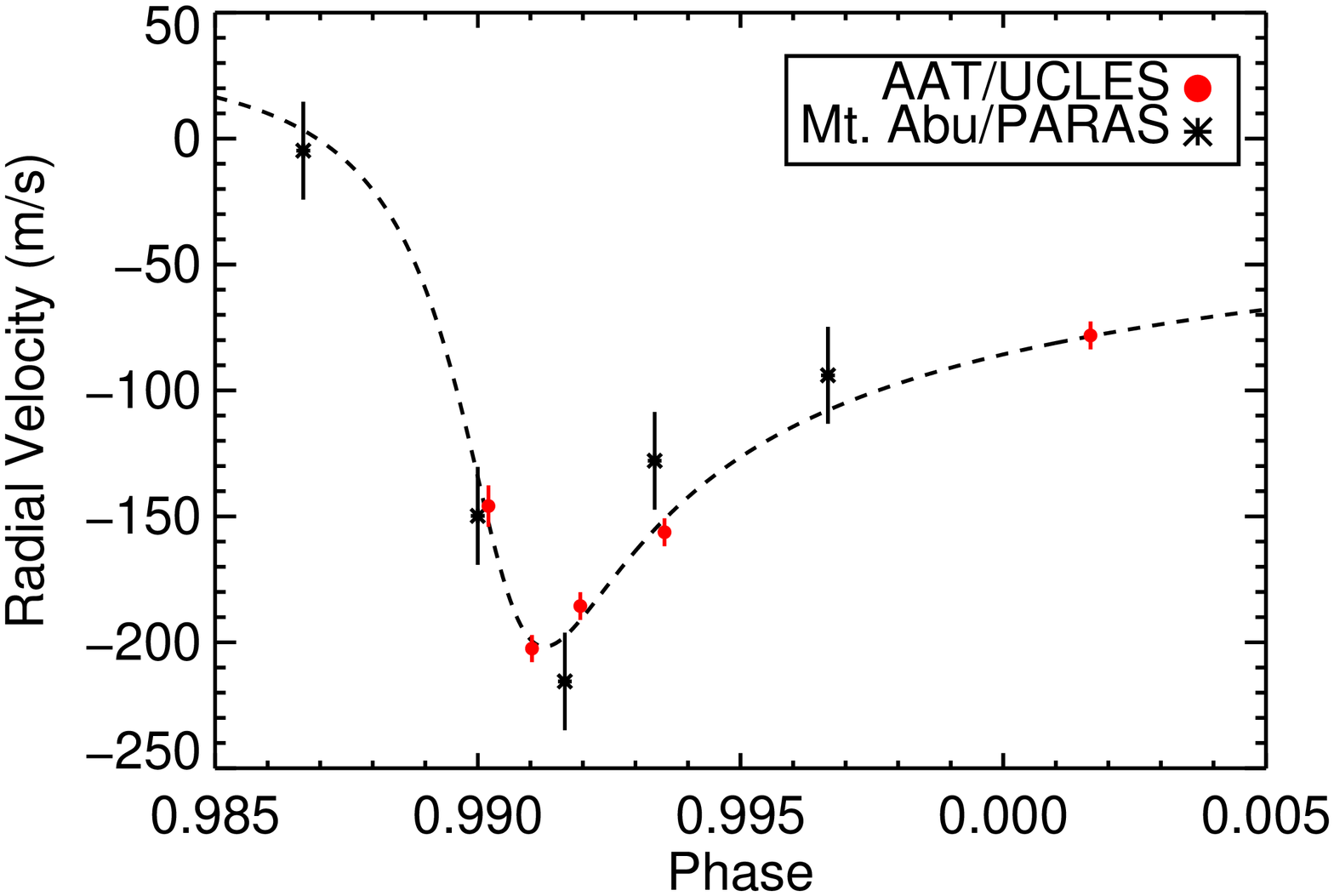}
  \caption{Top: All 57 RV measurements from AAT/PARAS observations of
    HD~20782 (see Tables \ref{rvs_aat} and \ref{rvs_paras}) along with
    the best-fit orbital solution (Table \ref{planet}). RV offsets
    between datasets have been accounted for in this figure. Middle:
    The RV data phased on the orbital the solution from
    Table~\ref{planet}, where phase zero corresponds to superior
    conjunction. Bottom: A zoomed version of the phased middle plot
    which shows the coverage during periastron passage.}
  \label{rv}
\end{figure}

We conclude that there is significant instrumental/observational
systematic noise in the PARAS data due to air mass, of order
20~m\,s$^{-1}$. We also examined the inclusion of a linear RV trend in
our model but found that this does not improve the quality of the
fit. The final orbital solution from the data is shown in
Table~\ref{planet}, where we have included the solution that uses the
AAT data only for comparison. The AAT+PARAS orbital solution includes
an offset between the AAT and PARAS datasets as a free parameter,
found to be $276.5 \pm 8.7$~m\,s$^{-1}$. The $\gamma$ parameter shown
in Table~\ref{planet} is the systemic velocity of the system with
respect to the zero point of the extracted RVs (relative to the
template spectrum). Thus, there is an offset between the $\gamma$
value reported in Table~\ref{planet} and the true systemic velocity,
reported by \citet{val05} to be 40.7~km\,s$^{-1}$. Our final AAT+PARAS
orbital solution is depicted in Figure~\ref{rv}. The bottom panel of
Figure~\ref{rv} shows the quality of the combined data coverage during
periastron passage for this highly eccentric planet, particularly the
additional coverage provided by the PARAS data.

We note that the transit time we calculate is sensitive to the weights
assigned to the PARAS and AAT data. The PARAS data favor a transit
time that is 0.2--0.3 days later than the AAT data. Because we do not
fully understand the source of the very large scatter in the PARAS
data, we should not assume that our errors are Gaussian. Fortunately,
BOOTTRAN uses bootstrapping to determine its parameter uncertainties,
which is appropriate for non-normally distributed residuals (although
the underlying fitter minimizes $\chi^2$, and so does assume Gaussian
errors).


\section{Astrometric Constraints on the Orbit}
\label{astrometry}

To constrain the inclination of the system and possibly refine the
estimate of the companion mass, we combine Hipparcos astrometry of
HD~20782 with the orbital parameters obtained from the radial velocity
observations. We use the new reduction of the Hipparcos data
\citep{van07a}, which presents a significant improvement in the
overall reliability of astrometric information \citep{van07b} and
includes the Intermediate Astrometric Data (IAD) product in a format
that facilitates the quest for signatures of orbital motion. Following
the method prescribed by \citet{sah11}, we use the spectroscopic
elements derived from our RV solution (Table~\ref{planet}) to search
for an orbital signature.

The 5 standard astrometric parameters for the Hipparcos solution of a
HD20782 can be obtained from the VizieR Catalogue \citep{van07a};
these are right ascension (RA, $\alpha$=50.01$^{\circ}$), declination
(dec, $\delta$=-28.85$^{\circ}$), proper motion in RA
($\mu_{\alpha}$=349.33 mas/yr) and dec ($\mu_{\delta}$=-65.92 mas/yr),
and parallax ($\varpi$=28.15 mas). The 5 spectroscopic parameters
required from the radial velocity analysis are period ($P$),
eccentricity ($e$), semi-amplitude ($K$), time of periastron ($T_0$),
and argument of periastron ($\omega$). Each Hipparcos observation is
reconstructed from the IAD and fit with a comprehensive model based on
12 parameters: the 5 standard astrometric parameters, the 5
spectroscopically derived parameters, the inclination ($i$), and the
longitude of the ascending node ($\Omega$). In practice, the
spectroscopic parameters are treated as constants since they are
considered reliable, and we work with 7 free parameters. The details
of the procedure are carefully described by \citet{sah11}, and we
follow their methods to calculate inclination, a new orbit, and the
significance of the orbit via the permutation test.

We begin by constructing a two-dimensional $i-\Omega$ grid, where we
solve for the remaining five parameters of the 7-parameter model, and
the corresponding $\chi^2$. The parameter values identified by the
minimum $\chi^2$ value are used as the starting point for an $AMOEBA$
minimization, using the downhill simplex method, to supersede the
limitations imposed by the resolution of the $i-\Omega$ grid. We then
perform 100,000 Monte Carlo simulations, where we generate 1,000 sets
of Hipparcos measurements from the existing data. For each set of
Hipparcos abscissae, we execute 100 random draws from the
spectroscopic parameters, in order to inculcate their uncertainties
into our result.  Each spectroscopic parameter is assumed to be a
gaussian distribution, with the RV solution and its errors serving as
the mean and standard deviation respectively. The Monte Carlo models
are then solved as described above, to produce 100,000 sets of
solution parameters.  The final parameters are defined as the median
of the associated distribution, while the errors are the interval
between the 15.85 percentile and the 84.15 percentile, encompassing
68.3\% of the values around the median.

For completeness, we report here our full set of final parameters, as
offsets to the Hipparcos values.  The changes in right ascension,
declination, parallax, proper motion in RA and declination, and the
argument of periastron are: $\Delta\alpha=0.3^{+1.4}_{-1.2}$ mas,
$\Delta\delta=1.0^{+1.3}_{-1.1}$ mas, $\Delta\varpi=0.2^{+0.7}_{-0.7}$
mas, $\Delta\mu_{\alpha^{\star}}=0.0^{+0.4}_{-0.4}$ mas/yr,
$\Delta\mu_{\delta}=0.1^{+0.6}_{-0.6}$ mas/yr.  We find an inclination
of ${2.7^{+2.3}_{-1.2}}^{\circ}$ and an argument of periastron
${202.5^{+59.3}_{-66.3}}^{\circ}$, but the solution is very poorly
constrained.

\begin{figure}
  \includegraphics[width=8.2cm]{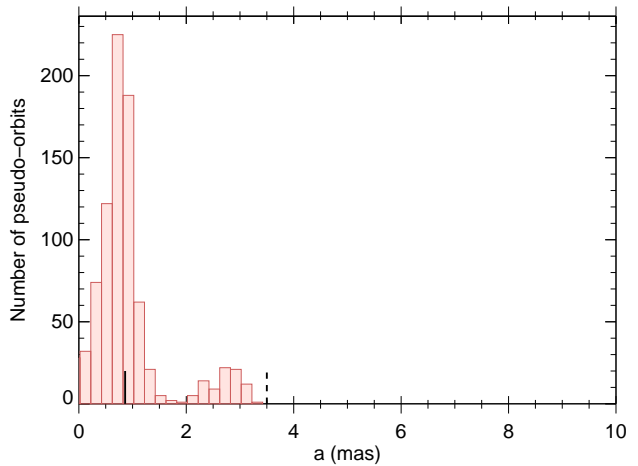}
  \caption{Histogram of the semi-major axes for 1000 randomly permuted
    pseudo-orbits of HD~20782. The pseudo-orbits are used to calculate
    the significance of the new orbit via the permutation test, as
    described by \citet{sah11}. The solid black line shows the actual
    best-fit solution and the dashed line represents the median
    Hipparcos single-measurement precision for this system.}
  \label{fig:hist}
\end{figure}

\begin{figure}
  \includegraphics[width=8.2cm]{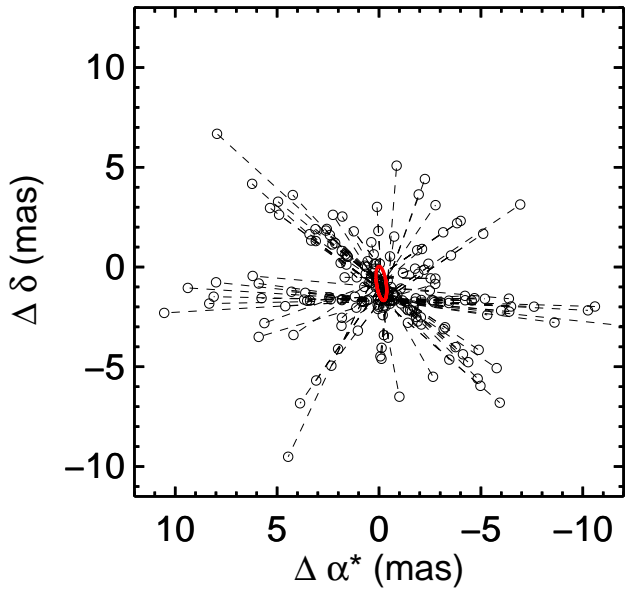}
  \caption{The red line shows the orbital signature detected in the
    Hipparcos data when combined with orbital parameters from the
    radial velocity solution. As projected on sky, North is up and
    East is left. Open circles mark the individual Hipparcos
    measurements. Dashed lines with orientation along the scan angle
    $\psi$ and length given by the residual of the orbital solution
    connect the measurements with the predicted location from our
    model. This illustrates the difficulty of detecting an orbit with
    such a small projected semi-major axis, given the median Hipparcos
    single-measurement precision on this target.}
 \label{fig:orbit}
\end{figure}

The astrometric data covers approximately two orbits for this system,
so phase coverage should not inhibit the recovery of any significant
orbital signatures. Unfortunately, the projected minimum semi-major
axis of our new solution is very small ($a \sin i = 0.05$~mas)
compared to the median Hipparcos single-measurement precision for this
target (3.5 mas), as shown in Figures \ref{fig:hist} and
\ref{fig:orbit}. We perform a permutation test to verify the
significance of our result by comparing the semimajor axis of the new
solution orbit with 1,000 pseudo orbits generated from random
permutations of the astrometric data, similar to \citet{sah11}. Figure
\ref{fig:hist} illustrates the calculation of a low significance orbit
(68.2\%, which is almost exactly at the 1$\sigma$ level of detection),
confirming that the Hipparcos data contains little or no orbital
signature. The new Hipparcos reduction has this target flagged as
having a good fit with just the 5 astrometric parameters, which is
consistent with the fact that adding the 7 RV parameters does not seem
to change the solution. \citet{sah11} emphasize that orbital solutions
at this significance level are prone to very large biases, and the
calculated values and their errors should be considered highly
suspect. We present our full set of orbital parameters here only to
facilitate future comparison of analytical methods, and not for direct
application.

\begin{figure}
  \includegraphics[angle=270,width=8.2cm]{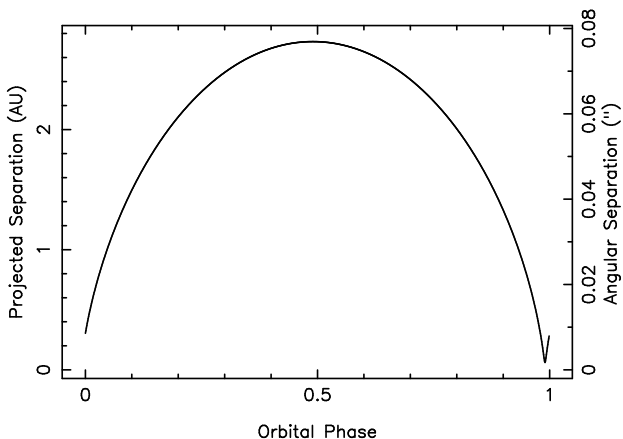}
  \caption{The projected (AU) and angular ($\arcsec$) separation of
    HD~20782b from the host star as a function of orbital phase, where
    phase zero corresponds to superior conjunction.}
  \label{sep}
\end{figure}

On the other hand, simulations by \citet{sah11} show that orbits are
always detected at the 3$\sigma$ level when the semi-major axis is at
least 70\% of the Hipparcos precision on a target. Any orbital
signature above 2.45 mas would have been detectable in this Hipparcos
dataset, and this helps to set an upper limit on the companion
mass. Using this assertion, we get a lower limit on inclination
(1.22$^{\circ}$) and an upper limit on the companion mass
(66~$M_J$). Although the consideration of astrometric data does not
allow us to put tight constraints on the inclination of the system,
the non-detection of an orbit allows us to rule out a stellar binary
system. Verification of this could be achieved through high-contrast
adaptive-optics imaging of the system at predicted apastron
passage. Figure~\ref{sep} shows the projected and angular separation
of the planet and star for one complete face-on orbit, where phase
zero corresponds to superior conjuction as described by
\citet{kan13}. An additional consequence of our astrometric constraint
is that the transit probability is increased by a small amount since
inclinations below 1.22 degrees are ruled out.

\section{Planetary Transit Prospects}
\label{transit}

As described in Section~\ref{science}, one of the most interesting
aspects of HD~20782b is the relatively large transit probability
compared with the orbital period. The transit probability is a
function of the stellar and planetary radii and the star--planet
separation along the line of sight \citep{kan08}. For HD~20782, we use
the stellar radius shown in Table~\ref{stellar} and adopt a planetary
radius of $R_p = 1.0$~$R_J$ given the minimum mass of 1.41~$M_J$ (see
Table~\ref{planet}) and using the mass-radius relationship described
by \citet{kan12a}.

If the planet were in a circular orbit with the same semi-major axis,
the transit probability would be 0.4\%. The extreme eccentricity of
the orbit results in star--planet separation of 0.061~AU at periastron
and 0.076~AU at inferior conjunction where a transit is possible. Such
a separation is similar to that of a hot Jupiter in a circular
orbit. Inferior conjunction occurs when $\omega + f = 90\degr$; in
this case, the true anomaly is $f = 307.9\degr$ at the time of
mid-transit. This orbital orientation results in an enhanced transit
probability of 7.1\%.

A further influence of the high eccentricity on the transit parameters
is the expected transit duration. Since the separation at inferior
conjunction is comparable to a hot Jupiter, the duration is likewise
reduced and has an amplitude of 0.13~days (3.1~hours) for a central
transit. The epoch of mid-transit shown in Table~\ref{planet} was
calculated using the same Monte-Carlo bootstrap method used to
calculate the orbital parameter uncertainties. The time of mid-transit
corresponds to a calendar date of 15 January 2015 and a UT of
7:02. The uncertainty on this time is 0.1~days which results in a
total 1$\sigma$ transit window of 0.33~days (7.6~hours). The estimated
depth of the transit is 0.96\% and so should be readily observable in
typical millimag photometry. However, the infrequent occurrence of
such events (see Section~\ref{science}) motivated observations from
both ground and space.


\section{Photometric Observations}
\label{photometry}

The derived physical and orbital properties of HD~20782b described in
previous sections motivated photometric monitoring of the host star
for stellar variability and planetary transit/phase signatures. Here
we describe our photometric observations and results in detail.


\subsection{APT Photometry}

We collected a total of 191 nightly photometric observations of
HD~20782 during its 2013--14, 2014--15, and 2015--16 observing seasons
to search for stellar variability. The observations were acquired with
the T8 0.80~m automatic photoelectric telescope (APT), one of several
automated telescopes operated by Tennessee State University (TSU)
located at Fairborn Observatory in southern Arizona. The T8 APT is
equipped with a two-channel precision photometer that uses a dichroic
filter and two EMI 9124QB bi-alkali photomultiplier tubes to measure
the Str\"omgren $b$ and $y$ pass bands simultaneously. We computed the
differential magnitudes of HD~20782 with respect to the mean
brightness of its three constant comparison stars. To improve the
precision further, we combined the differential $b$ and $y$
observations into a single $(b+y)/2$ ``passband.''  The TSU APTs,
their precision photometers, observing strategy, and data reduction
techniques are described in detail by \cite{hen99}. A summary of the
photometric data for HD~20782 is given in Table~\ref{aptsum}.

\begin{deluxetable}{ccccc}
  \tablecaption{\label{aptsum} Summary of photometric observations for
    HD~20782}
  \tablewidth{0pt}
  \tablehead{
    \colhead{Observing} &
    \colhead{} &
    \colhead{Julian Date Range} & 
    \colhead{Mean} &
    \colhead{Sigma} \\
    \colhead{Season} &
    \colhead{$N_{obs}$} &
    \colhead{(HJD -- 2,400,000)} &
    \colhead{(mag)} &
    \colhead{(mag)}
  }
  \startdata
  2013--14 &  43 & 56622--56701 & 1.01241 & 0.00183 \\
  2014--15 &  89 & 56943--57045 & 1.01206 & 0.00228 \\
  2015--16 &  59 & 57293--57390 & 1.01128 & 0.00171
  \enddata
\end{deluxetable}

\begin{figure}
  \includegraphics[width=8.2cm]{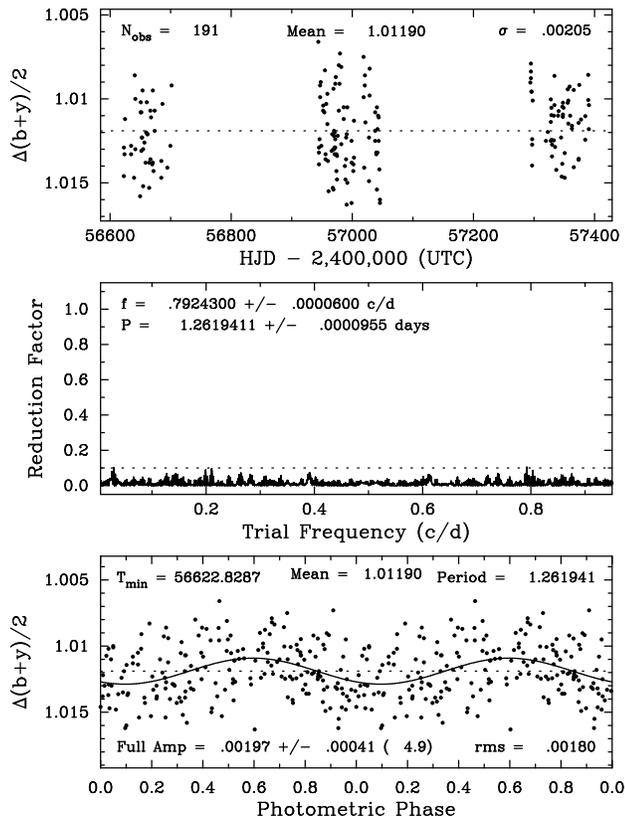}
  \caption{HD~20782 APT photometry acquired during three consecutive
    observing seasons. Top: The relative photmetry as a function of
    Heliocentric Julian Date. Middle: The power spectra from a fourier
    analysis of all seasons photometry. Bottom: Sinusoidal fit to the
    most significant period found from the fourier analysis. Our
    analysis described in the text demonstrates that this period is
    spurious.}
  \label{aptphot}
\end{figure}

The nightly observations of HD~20782 are plotted in the top panel of
Figure~\ref{aptphot} in our $(b+y)/2$ passband. The observing seasons
are quite short from Fairborn Observatory, only three months in
length, because of the star's southerly declination of $-29\arcdeg$.
The observations scatter about their grand mean (indicated by the
horizontal dotted line) with a standard deviation of 0.00205 mag, as
given in the upper right corner of the top panel. This is essentially
the limit of precision for the HD~20782 observations because the
star's southerly declination results in all measurements being made at
air mass values between 2.0 and 2.4 (see \cite{hen99}, Figure~8).

The middle and bottom panels of Figure~\ref{aptphot} show the
frequency spectrum of the data set and the phase curve computed with
the best frequency, respectively. Our frequency analysis is based on
least-squares sine fits with trial frequencies between 0.01 and 0.95
c/d, corresponding to periods between one and 100 days. The goodness
of fit at each frequency is measured as the reduction factor in the
variance of the original data, whose value lies between the extremes
of 0.0 and 1.0. A reduction factor of 0.0 corresponds to the case
where the variance in the residuals from a least-squares sine fit to
the observational data at some trial frequency have the same value as
the variance in the original data, i.e., no reduction in the variance
takes place at that particular frequency. A reduction factor of 1.0
corresponds to the extreme case where the variance in the residuals of
the sine fit is 0.0, i.e., the sine curve fits the data perfectly with
no residuals. The frequency spectrum in the middle panel shows several
peaks with reduction factors near 0.1, but no peak stands out above
the others to suggest a stellar rotation period.  We ran simulations
adding computed sine curves to our data sets and found that coherent
variations with peak-to-peak amplitudes of $\sim0.004$ mag or larger
would be detectable in our light curves. This places an upper limit to
any periodic modulation for HD~20782, such as rotational modulation
caused by starspots. This is consistent with the low level of magnetic
activity (logR'$_{HK}$ = -4.91) given in the discovery paper of
\citet{jon06} and demonstrates that our best-fit period of 1.2619 days
in the bottom panel is spurious.  In addition to the absense of
rotational modulation, we find no evidence for longer-term
variability; the three seasonal means in Table~\ref{aptsum} scatter
about their grand mean with a standard deviation of only 0.00058 mag.


\subsection{MOST Observations and Transit Window}
\label{most}

\begin{figure*}
  \begin{center}
    \includegraphics[angle=270,width=15.5cm]{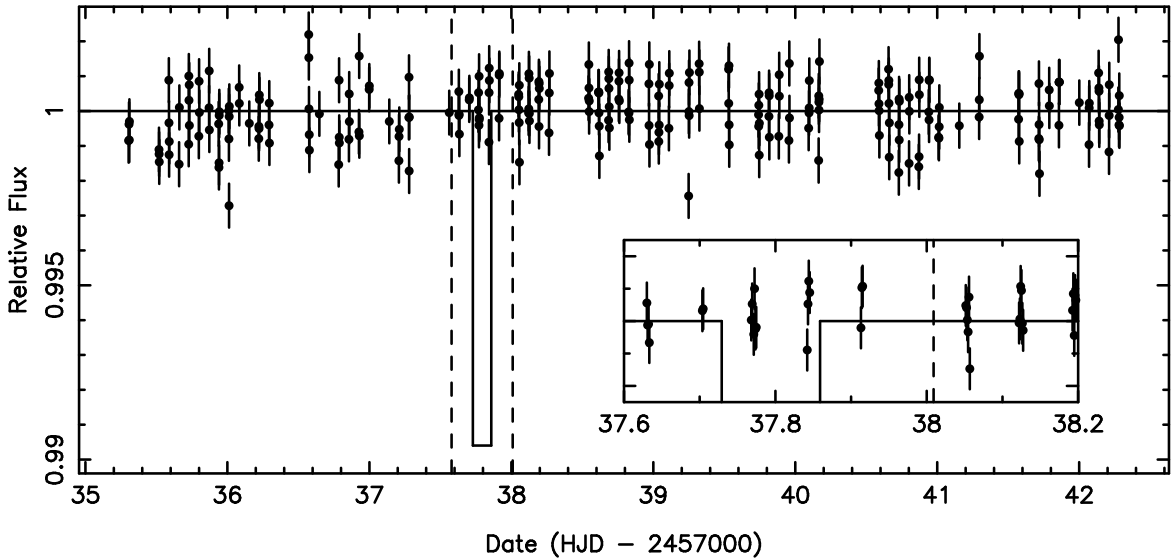}
  \end{center}
  \caption{MOST photometry of HD~20782 acquired for $\sim$7~days
    surrounding the predicted transit mid-point. The solid line
    indicates the location and depth of a possible transit and the
    vertical dashed lines are the boundaries of the 3$\sigma$ transit
    window.}
  \label{transitfig}
\end{figure*}

Given the size of the transit windows and the relatively infrequent
opportunities to observe them (see Sections \ref{science} and
\ref{transit}), we elected to make use of the Microvariability and
Oscillations of STars (MOST) satellite to observe HD~20782 during the
next scheduled transit window. MOST has an aperture of 15~cm and a
filter passband covering the range 375--675~nm, making it well suited
to obtain precision optical photometry of very bright stars
\citep{wal03,mat04}.

Observations of HD~20782 commenced at HJD 2457035.3 (2015 January 12
19:11 UT) and concluded 7 days later at HJD 2457042.3 (2015 January 19
19:11 UT). The predicted time of mid-transit (see Table~\ref{planet})
was BJD 2457037.794 (2015 January 15 07:02 UT). The star is outside of
MOST's Continuous Viewing Zone and so required observations outside of
normal operational parameters. For each 101~min orbit, MOST was able
to acquire the target field for 20~mins. Exposure times were 0.6~secs
to allow for both the brightness of the target and scattered light due
to the roll angle of the spacecraft with respect to the Sun. This
resulted in photometry with a 1$\sigma$ RMS precision of 0.07\%.

During the week of MOST observations, a total of 257 measurements of
HD~20782 were acquired. The resulting relative photometry of the data
are shown in Figure~\ref{transitfig}, along with a solid line that
indicates the predicted location and depth of a possible transit. The
1$\sigma$ transit window (0.33~days) was described in
Section~\ref{transit}. We use the 3$\sigma$ transit window (0.43~days)
to draw vertical dashed lines in Figure~\ref{transitfig}. A central
transit of the planet (impact parameter of $b = 0$) is ruled out for
most locations within the transit window. The cadence of the
observations is such that a transit duration half that of a central
transit could have been missed within the 1$\sigma$ transit
window. Such a duration corresponds to an impact parameter of $b =
0.87$, above which transits cannot be ruled out by our photometry.

A further consideration is the detection of the Rossiter-McLaughlin
(R-M) effect during a possible transit. The amplitude of the R-M
effect is shown by \citep{gau07} to be
\begin{equation}
  K_R = v \sin i \frac{(R_p/R_\star)^2}{1 - (R_p/R_\star)^2}
  \label{rm}
\end{equation}
Using our stellar parameters from Table~\ref{stellar} and the transit
parameters described in Section~\ref{transit}, the amplitude of the
R-M effect for a transit of HD~20782b is predicted to be
$\sim$15.5~m\,s$^{-1}$. Two of our RV measurements (one each from AAT
and PARAS) are within the transit window, shown close to 0.5 phase in
the bottom panel of Figure~\ref{rv}. Neither of these measurements
show evidence of any significant deviation from our Keplerian
model. Thus the RV data are consistent with the MOST photometry
leading to the conclusion that the planet does not transit the host
star.


\subsection{Evidence of Phase Variations}
\label{phase}

The phase variations of a planet as it orbits the host star has become
a detectable signature in the era of high-precision
photometry. Numerous examples of phase signatures have been detected
from the planets detected with the {\it Kepler} mission
\citep{est13,est15}.

Exoplanets that are close to their host stars have generally been
found to have low geometric albedos, such as the low geometric albedo
of HAT-P-7b \citep{wel10} and the null detection of phase variations
from HD~209458b \citep{row08}. There are exceptions to the rule,
however, such as the case of Kepler-7b \citep{dem11}, and it is likely
that a greater understanding of atmospheric processes is needed to
explain this diversity \citep{dem14}. \citet{kan10} developed a
geometric albedo model that scales the geometric albedo with
star--planet separation. The implication of the model for planets in
eccentric orbits is that the geometric albedo is time dependent, with
an assumption that reflective/scattering condensates in the upper
atmosphere are removed during periastron passage by the increase in
radiative flux from the host star. The generalized expression for the
planet to host flux ratio is given by
\begin{equation}
  \epsilon(\alpha,\lambda) \equiv
  \frac{f_p(\alpha,\lambda)}{f_\star(\lambda)} = A_g(\lambda)
  g(\alpha,\lambda) \frac{R_p^2}{r^2}
  \label{fluxratio}
\end{equation}
where $\alpha$ is the phase angle, $A_g$ is the geometric albedo, $g$
is the phase function, $R_p$ is the planetary radius, and $r$ is the
star--planet separation. This separation is given by
\begin{equation}
  r = \frac{a (1 - e^2)}{1 + e \cos f}
  \label{separation}
\end{equation}
where $f$ is the true anomaly. The phase angle, $\alpha$, is defined
to be zero at superior conjunction. A model of geometric albedo
time-dependence assumes that the planetary atmosphere responds to the
change in incident flux on timescales comparable to the duration of
the periastron encounter. This effect has been modeled for the
eccentric planet HAT-P-2b at infrared wavelengths by
\citet{lew13}. Thus \citet{kan10} predicted that, although the largest
phase variations of eccentric planets occur during a relatively short
fraction of their orbital phase, the amplitude of the signature would
be lowered by the subsequent darkening of their atmospheres during
periastron.

\begin{figure*}
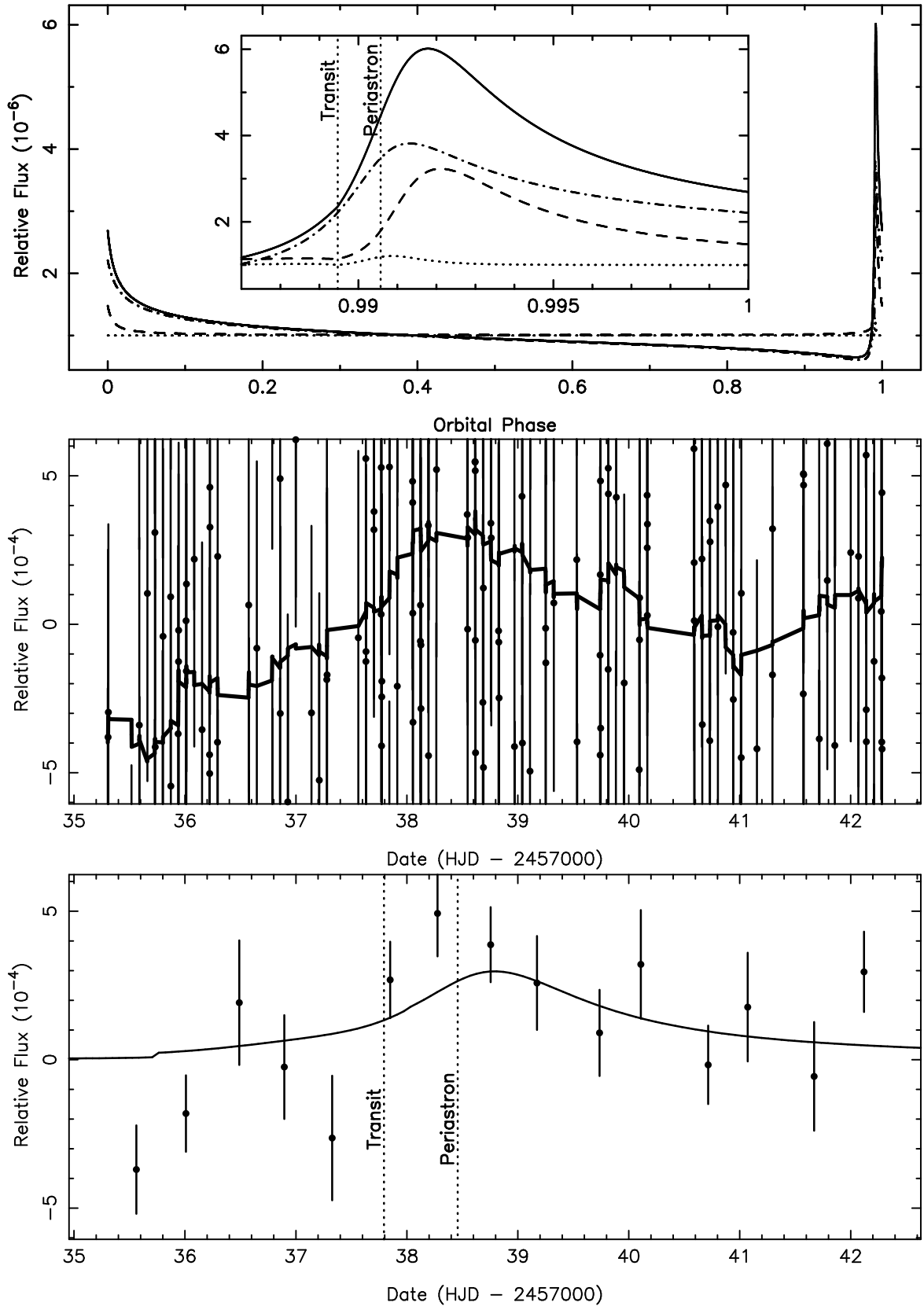

  \begin{center}
    \includegraphics[angle=270,width=15.5cm]{f08a.ps} \\
    \includegraphics[angle=270,width=15.5cm]{f08b.ps} \\
    \includegraphics[angle=270,width=15.5cm]{f08c.ps} \\
  \end{center}
  \caption{Top: The predicted flux variations of the HD~20782 system
    due to reflected light from the planet (dashed line), ellipsoidal
    variations (dotted line), and Doppler boosting (dot-dashed
    line). The sum of these three effects is shown as a solid
    line. This assumes a time-varying geometric albedo, as formulated
    by \citet{kan10}. The zoomed panel shows the maximum phase
    variation along with the orbital phase location of periastron and
    the predicted transit time described in Section
    \ref{transit}. Middle: The MOST data with the running average
    shown as a solid line. Bottom: The binned MOST data along with a
    model of the phase variations.}
  \label{phasefig}
\end{figure*}

For HD~20782b, we calculated the predicted phase variations of the
planet with respect to the inferior conjunction (transit) and
periastron times, shown as a dashed line in the top panel of
Figure~\ref{phasefig}. These orbital locations are very close to each
other (see Figure~\ref{orbitfig}), separated by only 0.66~days. The
location of superior conjunction where $\alpha = 0$ occurs 5.63 days
after the periastron passage. All three of these orbital locations are
covered by the MOST observations described in Section~\ref{most}. We
include the additional effects of ellipsoidal variations
\citep{dra03,zuc07,kan12b} and Doppler boosting \citep{loe03,fai11} in
the top panel of Figure~\ref{phasefig}, shown as dotted and dot-dashed
lines respectively. The combined effect of all three (phase
variations, ellipsoidal variations, and Doppler boosting) is shown as
a solid line. For the ellipsoidal component, we have assumed a gravity
darkening exponent of $\beta = 0.32$ \citep{luc67}. For the Doppler
boosting coefficient, we calculate a value of $\alpha_{\mathrm{beam}}
= -1.21$ using the stellar temperature from Table~\ref{stellar} and
the methodology of \citet{loe03}. Using the model of a
distance-dependent geometric albedo and Hilton phase function
\citep{kan10}, we determined that the amplitude of the phase
variations is comparable to the Doppler boosting, whereas the
ellipsoidal component is a minor contribution to the total flux
variations. Another point worth noting is that this model assumes an
orbit that is close to edge-on. The effect of orbital inclination on
the relative amplitudes of the three contributing components is minor
except for orbits close to face-on \citep{kan11a}.

As described in Section~\ref{most}, the original intent of acquiring
the MOST data was for the purpose of observing a potential transit
event. Evidence of phase variations was unexpected due to the low
predicted amplitude shown in the top panel of
Figure~\ref{phasefig}. To determine the overall trend in the MOST
data, we calculated a running mean of the data using 20 data points
either side of each measurement to calculate the running mean at that
location. The results of this calculation are shown as a solid line
along with the individual measurements (including error bars) in the
middle panel of Figure~\ref{phasefig}, where we have adjusted the
vertical scale of the plot to the range of the running mean values,
using the average of the running mean values as the zero-point. The
apparent brightening of the host star between the truncated dates of
38 and 39 on the plot is where the peak of the phase variations are
predicted to occur. This was diagnosed from an instrumentation point
of view, and it was determined that the change in the brightness was
not caused by any aspect of the MOST instrumentation or data reduction
issues.

We tested the likelihood whether this could be caused by an alignment
between intrinsic stellar variability and the expected periastron
passage, by conducting a Monte-Carlo simulation in which we treat the
observed data as representative of possible stellar variability and
randomly rearrange the data to see how often a similar chance
alignment can occur. Each random permutation of the observed flux
values to the times of observation resulted in a new dataset for which
the running mean was calculated and then analyzed for significant
peaks in the flux. The percentage of simulations for which a specific
criteria was met was taken as the probability that the criteria would
have been satisfied by chance. Based on 10,000 realizations of this
simulation, the probability of a peak occurring in the 38--39 date
range is $\sim$17\%, and the probability of that peak being of equal
or greater amplitude than the observed peak is $\sim$4\%. If indeed
the observed peak is related to the close passage of the planet to the
star, the flux variations may indicate that the assumption by
\citet{kan10} that the presence of reflective condensates in a
planetary atmosphere changes on timescales comparable with the
periastron passage is likely incorrect for highly eccentric orbits. In
fact, the larger the eccentricity, the more inconsistent the
assumption becomes with the radiative and advective timescales of the
atmosphere. Furthermore, the possible presence of phase variations
indicates the companion is not self-luminous, further supporting the
claim that the companion is planetary rather than stellar in nature
\citep{kan12b}.

To investigate this further, we binned the MOST photometry into 15
evenly spaced time intervals. The best-fit model to the binned data is
shown in the bottom panel of Figure~\ref{phasefig} where the model
includes ellipsoidal variations and Doppler boosting as well as phase
variations. The best-fit inclination of the planetary orbit is $i =
30\degr$. A fit of the data to both the described model and a constant
model resulted in $\Delta \chi^2 = 24$ which shows that the phase
model is quantitatively favored. However, the model uses a companion
radius of $\sim$5 Jupiter radii and a geometric albedo of unity, which
is a physically unlikely scenario. Thus there is either an additional
component missing in our model of the data, or the data may be
insufficient to fully characterize the flux variations, or some
combination of the two. As noted above, the most compelling aspects of
the variations described here are the timing of the variations with
those predicted by the phase model, combined with the extreme
eccentricity of the planet. This system is clearly highly unusual
amongst the known exoplanets, and we cannot exclude the possibility of
unaccounted for physics occuring during the extreme conditions of
periastron.

\begin{figure}
  \includegraphics[angle=270,width=8.2cm]{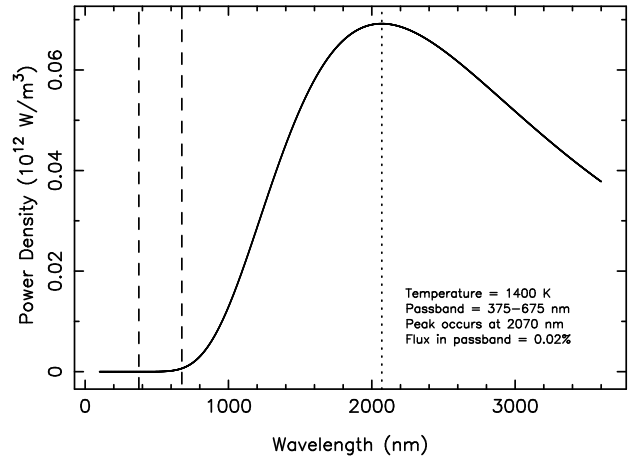}
  \caption{The predicted blackbody flux of HD~20782b, assuming a
    calculated temperature of $\sim$1400~K. The passband boundaries of
    MOST are indicated by the vertical dashed lines. The blackbody
    calculation assumes zero Bond albedo and zero heat redistribution
    (hot dayside model) and thus represents a maximum flux
    scenario. Of the integrated flux, 0.02\% falls within the MOST
    passband.}
  \label{blackbody}
\end{figure}

A possible missing factor is that of thermal emission. This has been
shown to be a significant component at the {\it Kepler} passband
\citep{dem11}. The {\it Kepler} passband however is significantly
broader than that used by MOST (see Section~\ref{most}). We calculated
this component for our observations by estimating the equilibrium
temperature of the planet. To do so, we assumed the most extreme case
of zero heat redistribution (hot dayside) and zero Bond albedo
\citep{kan11b}. This produces a peak equilibrium temperature at
periastron of $\sim$1400~K. The resulting blackbody spectrum is shown
in Figure~\ref{blackbody} along with the passband of MOST, depicted as
vertical dashed lines. Of the integrated flux from the thermal
emission, only 0.02\% of the total flux falls within the passband of
our observations. This corresponds to a flux ratio of planet to star
thermal emission in the MOST passband of $\sim 1.5 \times 10^{-6}$. We
conclude that any phase variations due to the planet are dominated by
the optical component. Further data with higher precision are needed
to confirm the presence of the variations and constrain the reflective
properties of this fascinating planet as it passes through periastron.


\section{Conclusions}
\label{conclusions}

Exoplanets in eccentric orbits remain some of the most intriguing
discoveries of recent decades. Although the semi-amplitude of their RV
variations is systematically higher, the orbits of highly eccentric
planets are difficult to characterize due the rapid variation at
periastron passage. We have refit the orbit of HD~20782b, consistent
with it being the most extreme of these eccentricity cases and have
provided new stellar and planetary parameters. Our RV measurements
acquired during the brief duration of periastron passage allow a
detailed orbital ephemeris to be constructed, despite the relatively
long period of $\sim$18 months. Our analysis of the {\it Hipparcos}
astrometry for HD~20782 constrains the inclination sufficiently such
that the companion is likely to be planetary rather than stellar. The
uncertainties associated with our astrometric analysis leave open the
possibility that the companion lies within the brown dwarf mass
regime. It is expected that further astrometric data from the Gaia
mission will significantly improve these constraints
\citep{per14}. Even with a relatively high transit probability of
$\sim$7\%, we have shown that the planet does not transit the host
star.

The possible phase variations soon after periastron might be induced
in part by stellar light reflected off the planet's
atmosphere. Although our modeling is incomplete, if this hypothesis is
true then it raises interesting questions regarding the conditions to
which such an extreme orbit exposes the planet. In particular, the
effects of rotation rate and radiative/advective timescales on
atmospheric dynamics may be overwhelmed by the short yet intense
conditions that occur at the closest approach of the planet to the
star. It has been further noted by several studies that the brightest
region of the planet is shifted westward of the substellar point,
caused by a relatively cloudy western hemisphere
\citep{dem11,est15,hu15,shp15}. Additionally, it seems likely that,
although planets in short-period orbits tend to have relatively low
geometric albedos, long-period planets in eccentric orbits retain a
high geometric albedo during the periastron passage since the
atmosphere does not have time to respond to the change in incident
flux. The result of this is a higher than expected flux ratio of the
planet to the star at optical wavelengths. Thus, eccentric planets
present a particularly lucrative observing opportunity for the study
of planetary atmospheres, provided one is able to accurately predict
when the peak flux variations are expected to occur.

Further observations of this system at times close to inferior
conjunction are highly encouraged. The next two times of inferior
conjunction predicted from our ephemeris are BJD $2457634.859 \pm
0.123$ (2016 September 3 8:36 UT) and BJD $2458231.924 \pm 0.153$
(2018 April 23 10:10 UT). In each case, the subsequent superior
conjunction occurs $\sim$6.29 days after the inferior
conjunction. Matching these times to those when the target is most
visible is not trivial and the timescale of the periastron passage is
best suited to continuous space-based observations. Possibilities for
these upcoming windows would be a perfect use for upcoming missions
that are optimized for bright star observations, such as the
CHaracterizing ExOPLanet Satellite (CHEOPS). A deeper understanding of
the orbits and atmospheres of eccentric planets are key milestones
towards unlocking the origin and nature of these mysterious objects.


\section*{Acknowledgements}

The authors would like to thank the anonymous referee, whose comments
greatly improved the quality of the paper. S.R.K and
N.R.H. acknowledge financial support from the National Science
Foundation through grant AST-1109662. G.W.H. acknowledges long-term
support from Tennessee State University and the State of Tennessee
through its Centers of Excellence program. H.R.A.J acknowledges
support from STFC via grants ST/M001008/1 and Leverhulme Trust
RPG-2014-281. The authors thank the Gurushikar, Mt. Abu Observatory
Staff and the PARAS technical staff for the observations with
PARAS. The PARAS program is supported completely by the Physical
Research Laboratory, Dept. of Space, Govt. of India. The results
reported herein benefited from collaborations and/or information
exchange within NASA's Nexus for Exoplanet System Science (NExSS)
research coordination network sponsored by NASA's Science Mission
Directorate.


\end{document}